\documentstyle[12pt]{article}
\begin{document}

\begin{titlepage}

\begin{flushright}
{INRNE-TH/02/93}
\end{flushright}

\vskip 0.6truecm

\begin{center}
{\large \bf  A $U(1)$ Current Algebra Model Coupled to \\ $2D$-Gravity\\}
\end{center}

\vskip 1.0cm

\begin{center}
M. Stoilov$^{*}$ and R. Zaikov$^{\star\ddag}$ \\
\vskip 0.2cm
{\it Institute for Nuclear Research and Nuclear Energy \\
Boul.~Tsarigradsko chaussee 72, Sofia 1784, Bulgaria}

\end{center}

\vskip 2.3cm

\rm
We consider a simple model of a scalar field with $U(1)$
current algebra gauge symmetry coupled to  $2D$-gravity
in order to clarify the origin of Stuckelberg symmetry in
the $w_{\infty}$-gravity theory.
An analogous symmetry takes place in our model too.
The possible central extension of the complete symmetry algebra
and the corresponding critical dimension have been found.
The analysis of the Hamiltonian and the constraints
shows that the generators of the current algebra, the
reparametrization and the Stuckelberg symmetries are not independent.
The connection of  the model with $w_{1+\infty}-$ and
$W_{1+\infty}$-gravity is discussed.

\vfill

\begin{flushleft}
\rule{5.1 in}{.007 in}\\
$^{*}$ {\small Supported by BFSR under contract Ph-20-91}\\
$^{\star}$ {\small Supported by BFSR under contract Ph-11-91}\\
$^{\ddag}$ {\small Bitnet address: ZAIKOV@BGEARN \\ }
\end{flushleft}

\end{titlepage}

\baselineskip 0.65cm

\newpage

\section{Introduction}

There is a great interest in $W$-gravities at present \cite{Hull}.
These theories are generalizations of the $2D$-gravity in which
interactions between new gauge fields and currents of spin  greater
than $2$ have been added to the Lagrangian.
It is possible to consider the limit of infinitly many currents (so
called $w_{\infty}$- and $W_{\infty}$-gravity, or if a spin-$1$ current
is added also --- $w_{1+\infty}$- and $W_{1+\infty}$ gravity).
In the simplest realization of the $w$-gravity --- the chiral $w$-gravity ---
a set of additional Stuckelberg type symmetries takes place \cite{Bergshoeff}.
It is possible, using an infinite system of additional currents, to obtain
covariant formulation of the $w$-gravity \cite{Bergshoeff}
and in this formulation there are Stuckelberg symmetries too.
In order to clarify their origin we investigate a simple model with
local $U(1)$ current algebra symmetry coupled to $2D$-gravity.
(Similar task has been considered in Ref.~\cite{Polyakov}.)
The generalization of the model to $w_{1+\infty}$ is straightforward.
The model possesses Stuckelberg symmetry and its explicit form is known
not only in the chiral gauge, but also in the covariant formulation
without introduction of additional currents.
The Hamiltonian approach shows that the generators of the Virasoro
symmetry are built up as powers of the
constraint which generates the current algebra (the Sugawara construction).
This means that they are not independent and this is the reason for
the Stuckelberg symmetry to arise.
One can consider (on the same footing as it was done for Virasoro
algebra) higher powers of the constraints without difficulties and
to obtain $w_{1+\infty}$ generators.

In what follows we need some well known facts
about $2D$ free massless scalar field.
Its action
\begin{equation} \frac{1}{2} \int d^{2}x\;
\partial_{+}\varphi\partial_{-}\varphi
\end{equation}
is invariant (up to total divergences) under the following global symmetries.
\begin{eqnarray}
\delta_{+}\varphi & = & k_{+}\partial_{-}\varphi; \;\;
 \partial_{+}k_{+} = 0, \label{a} \\
\delta_{-}\varphi & = & k_{-}\partial_{+}\varphi; \;\;
 \partial_{-}k_{-} = 0, \label{b} \\
\delta_{3}\varphi & = & k_{3}; \;\; \partial_{+}\partial_{-}k_{3} = 0.
\label{c}
\end{eqnarray}
Symmetries (\ref{a}--\ref{b})  are respectively left and right
reparametrizations and symmetry
(\ref{c}) is the translation one ($U(1)$ if $x$ is a periodic field).
It is possible to make symmetries (\ref{a}--\ref{c}) local (i.e. with
arbitrary parameters) by adding suitable gauge fields coupled to the
corresponding  currents ($\partial_{\pm}\varphi\partial_{\pm}\varphi$ and
$\partial_{\pm}\varphi$).
For instance, the action with gauged left symmetries reads
\begin{equation}
\frac{1}{2}\int d^{2}x \;(\partial_{+}\varphi\partial_{-}\varphi -
A_{+}\partial_{-}\varphi - h_{++}\partial_{-}\varphi\partial_{-}\varphi).
\label{4}
\end{equation}
The same is the structure of the chiral $w$-gravity action.
It is known \cite{Bergshoeff} that in chiral $w$-gravity action a set of
additional Stuckelberg symmetries (which link gauge fields only) take place.
There is a local symmetry of the same kind in action (\ref{4}) also, namely
\begin{eqnarray}
\delta h_{++} & = & k \nonumber \\
\delta A_{+}  & = & -k\partial_{-}\varphi. \label{5}
\end{eqnarray}
This symmetry is a result of the fact that when one gauges symmetries
(\ref{a}-\ref{c}) they are not independent anymore: one can obtain the gauged
transformations (\ref{a}-\ref{b}) with a special choice of the parameter of
the gauged transformation (\ref{c}), i.e. in action (\ref{4}) there are more
gauge fields than real symmetries.
It is the Stuckelberg symmetry which reduces the gauge fields to the correct
number.
We note that the model with action (\ref{4}) is equivalent to the model of
chiral $w_{\infty}$-gravity with additional coupling to $U(1)$
current provided Stuckelberg symmetry in $w_{\infty}$-gravity is
fixed so, that gauge fields for currents with spin $ > 2$ are zero.

The connection with $w_{\infty}$-gravity is less trivial if the matter field
takes value in some nonabelian Lie algebra $\cal G$ (e.g. $SU(N),
SL(N,R),$ etc.).
In this case the free Lagrangian is
\begin{equation}
\frac{1}{2}\int d^{2}x\;{\rm Tr}(\partial_{+}\Phi\partial_{-}\Phi) =
\frac{1}{2}\int d^{2}x\;\partial_{+}\Phi^{a}\partial_{-}\Phi^{a},
\label{22}   \end{equation}
where $\Phi \in {\cal G}$ and $a = 1,..,d; d = dim{\cal G}$.
Such Lagrangian has been used intensively as a starting point in
the construction of $W_{\infty}$-gravity \cite{Hull}, \cite{Pope}.
The basic idea in these considerations is to add to action (\ref{22})
a minimal coupling with properly realized currents of spin $\geq 2$.
When we add a coupling with spin-$1$ currents also (analogous to that
in action \ref{4}) the actions reads
\begin{equation}
\int d^{2}x \;\left\{tr(\frac{1}{2}\partial_{+}\Phi\partial_{-}\Phi
- A_{+}\partial_{-}\Phi) -
\sum_{j=0}^{\infty}h_{j}tr(\partial_{-}\Phi)^{j+2}\right\}.
\label{R7}
\end{equation}
We note that in action (\ref{R7}) the currents $\partial_{-}\Phi$
are not the $w_{1+\infty}$ spin-$1$ currents because
Tr$(\partial_{-}\Phi) \equiv 0$.
When $A_{+}=0$ action (\ref{R7}) coincides with the $w_{\infty}$-gravity
action considered in Ref.\cite{Hull,Bergshoeff}.
In the latter case only currents Tr$(\partial_{-}\Phi)^{j+2}$ for
$j=0,..,r-1,\; r = rank\cal G$ are independent.
The existing Stuckelberg symmetries permit to set all $h_{j} \equiv 0$
for $j\geq r$ (gauge fixing conditions).
However, when $A_{+} \not\equiv 0$, transformations
\begin{eqnarray}
\delta h_{j} & = & k_{j} \nonumber \\
\delta A_{+} & = & -\sum_{j=0}^{\infty}k_{j}(\partial_{-}\Phi)^{j+1}
\label{R8}
\end{eqnarray}
are also symmetries of action (\ref{R7}). Then, as in $U(1)$ case,
fixing Stuckelberg symmetries (\ref{R8}) by imposing
$h_{j} = 0 \;\;\forall j$ we are left with only one gauge field ---
$A_{+}$.
This gauge field takes value in the algebra $\cal G$ as the field
$\Phi$ does, while gauge fields $h_{j}$ are isotopic scalars.
The latter shows that action (\ref{R7}) differs from the action for
$w_{1+\infty}$-gravity where the gauge field for the spin-$1$
current is $\bf C$ valued also.
\section{Covariant formulation}

A natural question arises in connection with the Stuckelberg symmetry
(\ref{5}):
Is there an analogous transformation between $A_{-}$ and $h_{--}$ in
the nonchiral theory?
The problem is that one can regard on the action (\ref{4}) as obtained after
gauge fixing of two local symmetries (namely (\ref{b}) and (\ref{c})) in
the non-chiral theory, imposing conditions
\begin{equation} A_{-} = h_{--} = 0. \label{6} \end{equation}
In this gauge the Stuckelberg symmetry between $A_{-}$ and $h_{--}$ (if
exists) disappears, i.e. with two gauge conditions we have fixed three
independent symmetries.
In order to answer all these questions we consider the $2D$-gravity
action in Weyl gauge \cite{Pope}
\begin{equation}
h_{+-} = \frac{1}{2}(1 + h_{++}h_{--}). \label{7}
\end{equation}
In gauge (\ref{7}) $\det(h)$ is an exact square and the $2D$-gravity action
takes the form
\begin{equation} \label{8}
A = \frac{1}{2} \int d^{2}x\frac{1}{(1 - h_{++}h_{--})}
(\partial_{+}\varphi - h_{++}\partial_{-}\varphi)
(\partial_{-}\varphi - h_{--}\partial_{+}\varphi).
\end{equation}
This action is general reparametrization invariant \cite{Pope} as
the standard $2D$-gravity one is.
Moreover, one obtains exactly action (\ref{8}) in the Hamiltonian
approach to the Nambu-Gato string.
Gauging the translation symmetry of the action (\ref{8}) we obtain
\begin{equation}
A = \frac{1}{2} \int d^{2}x\frac{1}{(1 - h_{++}h_{--})}
                \nabla_{+}\varphi\nabla_{-}\varphi), \label{9}
\end{equation}
where $\nabla_{\pm}\varphi = \partial_{\pm}\varphi - A_{\pm} -
h_{\pm \pm}\partial_{\mp}\varphi + h_{\pm \pm}A_{\mp}$.
It turns out that action (\ref{9}) is invariant not only under local
translations
\begin{equation}
\delta\varphi = a, \;\; \delta A_{\pm} = \partial_{\pm}a \label{10}
\end{equation}
but also under an additional Stuckelberg-type symmetry
\begin{eqnarray}
        \delta\varphi & = & 0                \nonumber \\
\delta h_{\pm \pm} & = & k_{\pm\pm}       \label{11} \\
    \delta A_{\pm} & = & - \frac{\nabla_{\pm}\varphi}{1 - h_{++}h_{--}}
k_{\pm\pm}, \nonumber
\end{eqnarray}
provided
\begin{equation}
h_{--}k_{++} + h_{++}k_{--} = 0.  \label{12}
\end{equation}
One sees from eq.~(\ref{12}) that there is only one Stuckelberg symmetry
even in the non-chiral theory and that under it
\begin{equation}
\delta h_{+-} = 0,  \label{13}
\end{equation}
i.e., the Stuckelberg symmetry does not break Weyl gauge.

The generalization of the action (\ref{9}) for the nonabelian Lie algebra
valued matter field $\Phi$ is straightforward and reads
\begin{equation}
A' = \frac{1}{2} \int d^{2}x\frac{1}{(1 - h_{++}h_{--})}
                tr\left(\nabla_{+}\Phi\nabla_{-}\Phi)\right),
\label{R9}
\end{equation}
where, as in the $U(1)$ case, $\nabla_{\pm}\Phi =
\partial_{\pm}\Phi - A_{\pm} -
h_{\pm\pm}\partial_{\mp}\Phi + h_{\pm\pm}A_{\mp}$, but now the
gauge field $A_{\pm} \in \cal G$.
The gauge symmetries of action (\ref{R9}) are the same as those of
action (\ref{9}) with the only difference that the parameter of
the transformation (\ref{10}) is $\cal G$ valued now.
\section{Hamiltonian approach}
In what follows we shall consider a variant of chiral $2D$-gravity with
an additional local $U(1)$ current algebra symmetry with the following action.
\begin{eqnarray}
A & = & \frac{1}{2}\int d^{2}x\{\partial_{+}\varphi\partial_{-}\varphi
- A_{+}\partial_{-}\varphi
- A_{-}\partial_{+}\varphi +  A_{+}A_{-} -  \nonumber \\
& & - h_{++}(\partial_{-}\varphi)^{2} + 2h_{++}A_{-}\partial_{-}\varphi -
h_{++}A_{-}^{2}\} \label{15}
\end{eqnarray}
This action may be viewed as obtained from action (\ref{9}) after
imposing gauge $h_{--} = 0$.
The local symmetries of action (\ref{15}) are
\begin{eqnarray}
\delta_{\epsilon}\varphi & = & \epsilon\partial_{-}\varphi  \nonumber \\
 \delta_{\epsilon}h_{++} & = & \partial_{+}\epsilon -
\partial_{-}\epsilon h_{++} + \epsilon\partial_{-}h_{++}  \nonumber \\
  \delta_{\epsilon}A_{-} & = & \partial_{-}(A_{-}\epsilon)  \nonumber \\
  \delta_{\epsilon}A_{+} & = & \epsilon\partial_{-}A_{+} +
\partial_{+}\epsilon A_{-}                                \nonumber \\
            \delta_{a}'x & = & a                      \nonumber \\
       \delta_{a}'h_{++} & = & 0                  \label{16}  \\
      \delta_{a}'A_{\pm} & = & \partial_{\pm}a      \nonumber \\
           \delta_{k}''x & = & 0                       \nonumber \\
      \delta_{k}''h_{++} & = & k                         \nonumber \\
       \delta_{k}''A_{+} & = & - k(\partial_{-}x -A_{-})   \nonumber \\
       \delta_{k}''A_{-} & = & 0.                        \nonumber
\end{eqnarray}
It is more convenient to write down their algebra and possible
(nontrivial) central extensions using the Fourier components $L_{n}$,
$T_{n}$ and $S_{n}$ of the Virasoro ($\delta_{\epsilon}$), the current algebra
($\delta_{a}'$) and the Stuckelberg ($\delta_{k}''$) symmetries.
\begin{eqnarray}
\left[ L_{n}, L_{m} \right] & = & (n-m)L_{n+m} + cn^{3}\delta_{n+m}
\nonumber \\
\left[ T_{n}, L_{m} \right] & = & nT_{n+m}       \label{add} \\
\left[ S_{n}, L_{m} \right] & = & (n-m)S_{n+m} + en^{3}\delta_{n+m}
\nonumber \\
\left[ T_{n}, T_{m} \right] & = & n\delta_{n+m} , \nonumber
\end{eqnarray}
all other commutators vanishing.

We are looking for constraints which generate through their Poisson
bracket relations the transformations (\ref{16}).
There are three primary constraints in the model with action (\ref{15})
\begin{eqnarray}
\pi_{+} & \approx & 0 \nonumber \\
\pi_{-} & \approx & 0 \label{18} \\
\pi_{h} & \approx & 0, \nonumber
\end{eqnarray}
where $\pi_{\pm}$ and $\pi_{h}$ are the momenta canonically
conjugated to $A_{\pm}$ and $h_{++}$.
The momentum conjugated to $\varphi$ we denote by $p$.
For the Hamiltonian we obtain
\begin{eqnarray}
H & = & \frac{1}{2(1 - h_{++})}(p^{2} + \varphi'^{2} +
\frac{1}{4}(A_{+} - A_{-})^{2} + (A_{+} + A_{-})p -  \nonumber \\
 & & 2h_{++}p(\varphi' + A_{-}) + (A_{-} - A_{+})(1 + h_{++})\varphi').
\label{19}
\end{eqnarray}
 From the consistency conditions that the Poisson brackets between the
constraints and the Hamiltonian should be closed we get secondary conditions
\begin{eqnarray}
            p & \approx & 0 \nonumber \\
         \varphi'& \approx & 0 \label{20} \\
A_{+} - A_{-} & \approx & 0. \nonumber
\end{eqnarray}
The gauge transformations (\ref{16}) are generated by the following
combinations of the constraints (\ref{18}) and (\ref{20})
\begin{eqnarray}
(p - \varphi')^{2} & \approx & 0 \nonumber \\
      p - \varphi' & \approx & 0 \label{21} \\
\pi_{h} - \frac{1}{1-h_{++}}\pi_{+}(p-\varphi'+\frac{1}{2}(A_{+}-A_{-}))
             & \approx & 0 \nonumber
\end{eqnarray}
Strictly speaking among eqs.~(\ref{20}) there are second class constraints.
This is because for Bose fields the current algebra (its
central charge) arises on a classical level \cite{Goddard}.
We shall neglect this difficulty because we expect vanishing of
the central charge after switching on the ghost terms.

It is possible to consider a model with $\cal G$ valued matter field too.
In this case the action analogous to (\ref{15}) is
\begin{eqnarray}
A & = & \frac{1}{2}\int d^{2}x\{\partial_{+}\Phi^{a}\partial_{-}\Phi^{a}
- A_{+}^{a}\partial_{-}\Phi^{a} -
A_{-}^{a}\partial_{+}\Phi^{a} + A_{+}^{a}A_{-}^{a} -    \nonumber \\
  &   & - h_{++}(\partial_{-}\Phi)^{2} +
2h_{++}A_{-}^{a}\partial_{-}\Phi^{a} - h_{++}A_{-}^{2} \} \label{15+}
\end{eqnarray}
The symmetry algebra of this action differs from (\ref{add}) trought the
current algebra commutators only which now are
\begin{eqnarray}
\left[ T_{n}^{a},L_{m}\right]      & = &  nT_{n+m}^{a}   \nonumber \\
\left[ T_{n}^{a}, T_{m}^{b}\right] & = &  n\delta_{n+m}\delta^{ab}
\label{add2}
\end{eqnarray}
Note that commutation relations (\ref{add2}) are not the $\hat{\cal G}$
ones.
These are commutators of $d$ independent $U(1)$ transformations
(translations in the algebra parameters space).
Their generators are constraints
\begin{equation}
p^{a} - {\Phi^{a}}' \approx  0 , \label{21+} \\
\end{equation}
where $p^{a}$ are the momenta conugated to $\Phi^{a}$.
%
%
\section{Discussion}
Using BRST procedure \cite{BFV} one can formally obtain in $U(1)^{d}$
case (a trivial generalization of $U(1)$ case, discussed above)
that the Virasoro anomaly cancels if
$d = 54$ and normal ordering constant in $L_{0}$ equals two.
There is no way, however, to cancel $U(1)$ current algebra anomaly in
standard BRST procedure.
In order to handle this anomaly one should add to current algebra
generators $T_{n}$ the ghost terms $b_{n-k}c_{k}$, where $c_{n}, b_{n}$
are ghost-antighost pairs corresponding to $T_{n}$.

We want to stress again that all our results are valid both for
abelian and nonabelian valued matter fields.
The reason is that in the case of nonabelian algebra $\cal G$
the energy momentum tensor (and all currents with
spin $> 2$ which take place in action (\ref{R7})) is built up not
from currents, corresponding to the $\cal G$ symmetry, but from
$\partial_{\pm}\Phi^{a}$, which are translation ($U(1)^{d}$) currents
in the model.
Therefore, when we consider coupling with these spin-$1$ currents, we
are gauging not the $\cal G$ symmetry but the translation one and so, there
is no difference in our model between the $\cal G$ and the $U(1)^{d}$ cases.

Finally, we want to make a short comment about the connection with the
$W_{\infty}$-gravity in the $U(1)$ case.
It is known \cite{Berg2} that the quantization deforms $w_{\infty}$ to
$W_{\infty}$ symmetry.
This will not happen in our model if we generalize it to $w_{1+\infty}$.
The reason is that the current algebra anomaly vanishes and therefore there are
no nontrivial terms in the currents OPE.

\end{document}